\documentclass[aps,superscriptaddress,eqsecnum,nofootinbib,preprintnumbers]{revtex4}

\usepackage{graphicx,epsfig}
\usepackage{amssymb,amsmath}
\usepackage{subfigure}

\usepackage{relsize}
\usepackage[utf8]{inputenc}

\usepackage{xcolor}

\newcommand{\be}{\begin{equation}}
\newcommand{\ee}{\end{equation}}

\newcommand{\sch}{Sch\"ucking }

\newcommand{\ba}{\begin{eqnarray}}
\newcommand{\ea}{\end{eqnarray}}
\newcommand{\beg}{\begin{gather*}}
\newcommand{\eng}{\end{gather*}}

\newcommand{\p}{\partial}

\def\aj{\ref@jnl{AJ}}                   % Astronomical Journal
\def\actaa{\ref@jnl{Acta Astron.}}      % Acta Astronomica
\def\araa{\ref@jnl{ARA\&A}}             % Annual Review of Astron and Astrophys
\def\apj{\ref@jnl{ApJ}}                 % Astrophysical Journal
\def\apjl{\ref@jnl{ApJ}}                % Astrophysical Journal, Letters
\def\apjs{\ref@jnl{ApJS}}               % Astrophysical Journal, Supplement
\def\ao{\ref@jnl{Appl.~Opt.}}           % Applied Optics
\def\apss{\ref@jnl{Ap\&SS}}             % Astrophysics and Space Science
\def\aap{\ref@jnl{A\&A}}                % Astronomy and Astrophysics
\def\aapr{\ref@jnl{A\&A~Rev.}}          % Astronomy and Astrophysics Reviews
\def\aaps{\ref@jnl{A\&AS}}              % Astronomy and Astrophysics, Supplement
\def\azh{\ref@jnl{AZh}}                 % Astronomicheskii Zhurnal
\def\baas{\ref@jnl{BAAS}}               % Bulletin of the AAS

\begin{document}

%\maketitle

\title{Swiss-cheese cosmologies with variable $G$ and $\Lambda$ from the renormalization group}

%\date{\today}

\author{Fotios K. Anagnostopoulos}
\email{fotis-anagnostopoulos@hotmail.com}
\affiliation{Department of Informatics \& Telecommunications, 
University of the Peloponnese,  Karaiskaki 70, Tripoli 221 00, Greece}

\author{Alfio Bonanno}
\email{alfio.bonanno@inaf.it}
\affiliation{INAF, Osservatorio Astrofisico di Catania, Via S.Sofia 78, 95123 Catania, Italy}
\affiliation{INFN, Sezione di Catania, Via S. Sofia 72, 95123 Catania, Italy}

\author{Ayan Mitra}
\email{ayan.mitra@nu.edu.kz}
\affiliation{The Inter-University Centre for Astronomy and Astrophysics (IUCAA), Post Bag 4, Ganeshkhind,\\ Pune 411007, India}
\affiliation{Nazarbayev University, School of Engineering, Republic of Kazakhstan, 010000}
\affiliation{Kazakh-British Technical University, Tole bi 59, Almaty, Republic of Kazakhstan, 050000}

\author{Vasilios Zarikas}
\email{vzarikas@uth.gr}
\affiliation{University of Thessaly,Department of Mathematics, 35100 Lamia, Greece}
\affiliation{Nazarbayev University, School of Engineering, Republic of Kazakhstan, 010000}

\begin{abstract}

%A natural explanation of the recent passage to a cosmic acceleration era can be the existence of small anti-gravity sources in each galaxy and clusters of galaxies. In the context of Asymptotic Safety it is expected configurations of matter to be associated with a non zero cosmological constant and effective dark matter at the same time. 
%A Relativistic "sum" of all these anti-gravity sources with the help of a Swiss cheese model, 
%can generate a recent acceleration. The present work extends a previous study, allowing now both Newton constant
%$G(k)$ and cosmological constant $\Lambda(k)$ to be running couplings related to the length or time scale of the problem. 
%The analysis reveals that working with galaxy clusters there is a natural explanation of the coincidence problem. 
%The analysis have been done working both with a scaling for $k$ that uses the radial 
%proper distance of a cluster of galaxies and with an alternative scaling that uses the observational galaxy matter density. 
%Both approaches are able to provide recent cosmic acceleration with correct characteristics suggesting a consistent convincing supportive result. Furthermore, there is no fine tuning.
%The recent transition to a phase of cosmic acceleration can be explained by a uniform distribution of anti-gravity sources across the Universe which contributes to the observed vacuum energy.

A convincing explanation for the nature of the dark energy and dark matter is still missing. 
In recent works a RG-improved swiss-cheese cosmology with an evolving cosmological constant
dependent on the \sch radius has been proven to be a promising model to explain the 
observed cosmic acceleration. 
In this work we extend this model to consider the combined scaling of the 
Newton constant $G$ and the cosmological constant $\Lambda$ according to the
IR-fixed point hypothesis. We shall show that our model easily generates the observed recent passage from deceleration to acceleration without need of extra energy scales, exotic fields or fine tuning.
In order to check the generality of the concept,  two different scaling relations have been analysed and we proved that both are in very good agreement with $\Lambda$CDM cosmology. We also show that our model 
satisfies the observational local constraints on $\dot{G}/G$.
\end{abstract}
\maketitle

\section{Introduction}
Although the $\Lambda$CDM scenario reproduces rather well a large amount of observational data, it is important to stress that its two main ingredients,  Dark Matter (DM) \cite{1970ApJ...159..379R} 
and Dark Energy (DE) \cite{Spergel:2006hy}, do not have a direct physical explanation. 
In particular, the observed accelerated expansion of the Universe is caused by the DE component
a mechanism which introduces new difficulties in terms of fine-tuning or 
cosmic coincidences \cite{Straumann:1999ia} for the presence of a significant vacuum energy (cosmological constant).  

For this reason the possibility that the cosmological constant is dynamically generated as a low-energy quantum effect  has been recently investigated by various authors \cite{2017PhRvD..95j3504W,2019PhRvL.123m1302C}.
In particular in \cite{Bonanno:2001hi} a cosmology with a time dependent cosmological constant and Newton constant whose dynamics arises from an underlying renormalization group flow near an infrared  attractive fixed point has been proposed for the first time. 
Its phenomenological implications have been discussed in \cite{Bentivegna:2003rr} and further 
developed in \cite{2012NJPh...14b5008B}.

The IR-fixed point hypothesis originates from  the asymptotically safe (AS) scenario for quantum gravity 
\cite{c11,c12}. According to AS a consistent quantum theory of gravity can be realized 
nonperturbatively by defining the theory around a non-gaussian fixed point 
for the dimensionless Newton's constant 
\cite{Litim:2011cp,Nagy:2012ef, Bonanno:2017pkg,Percacci:2017fkn,Eichhorn:2018yfc,Pereira:2019dbn,Reuter:2019byg,Reichert:2020mja,Platania:2020lqb}.
In fact this mechanism is not qualitatively different from what occurs in the non-linear $\sigma$-model
in $d=3$: here too,  although the theory is not  perturbatively renormalizable, a continuum limit 
at non-vanishing value of the coupling constant can be defined beyond perturbation theory
\cite{Polyakov:1993tp,Codello:2008qq}. 

Assuming that the relevant scale of energy or momentum scales with the distance as the coarse-graining, 
``resolution" scale of the renormalization group,  the spatial or temporal evolution of the Newton's constant  and cosmological constant is dictated by the renomalization
group flow \cite{Platania:2020lqb}. The possibility of explaining DM in spiral galaxies as an infrared effect of the running 
of $G$ has been discussed in \cite{Reuter:2004nv,Reuter:2004nx}, 
while complete cosmological histories based on RG-trajectories emanating from the UV fixed point up to the IR regime have been discussed in
\cite{Reuter:2005kb,Bonanno:2007wg}. Non-singular cosmological models have appeared
in \cite{Kofinas:2016lcz,Bonanno:2017gji} and a mathematical 
formalism to couple the RG evolution to the Einstein equations have been discussed in 
\cite{Reuter:2003ca} and \cite{Bonanno:2020qfu}.

In \cite{Zarikas:2017gfv} a further step in the direction of explaining the DE content in the Universe with a running cosmological constant has been proposed.  The basic idea is to consider the contribution of an homogeneous distribution of  antigravity sources associated with the  matter content at galactic and cluster scale. A swiss-cheese cosmology represents an elegant mathematical framework to implement this idea
which has been elaborated also in \cite{sc1,sc2,sc3}. In this work we would like to extend the original RG-improved  swiss-cheese model to include the running of the Newton's constant according to the IR-fixed point mechanism.
In this case, the  scaling behavior of $G$ and $\Lambda$ is determined by the simple law
\begin{equation}
    G_k \equiv G(k)\sim g_\ast / k^2,   \quad \quad \Lambda_k \equiv \Lambda(k)\sim \lambda_\ast k^2 
\end{equation}
in the $k\rightarrow 0$ limit. This behavior could be the result of a ``tree-level"
renormalization induced by a non-zero positive cosmological constant in the IR, according to the
singular behavior of the $\beta$-functions in the IR \cite{Nagy:2012rn,Christiansen:2012rx,Biemans:2016rvp}.
The aim of this study is to further elaborate the IR-fixed point scaling within the swiss-cheese cosmology. In particular it will be shown that it is possible to consistently describe a phase of accelerated expansion of the universe without the introduction of a DE component.

The structure of this paper is the following: in section II the mathematical 
the swiss-cheese idea and the basic equations will be discussed. In Section III the IR fixed point
hypothesis and the RG evolution will be presented as a function of the \sch Radius and of the
matter density. In section IV the numerical results are discussed and section V is devoted to the
conclusions.

\section{Swiss cheese cosmological model}
\label{II}

The original Einstein-Strauss model or otherwise called swiss-cheese model, \cite{ES}, 
describes many homogeneously distributed black holes smoothly matched within a cosmological metric.
The key idea is to match a Schwarzshild spherical vacuole in a homogeneous isotropic cosmological spacetime. 
The analysis finally generates a dust Friedman-Robertson-Walker (FRW) cosmology assuming many Schwarzschild black holes as the matter content.

   	It is also possible to show that respecting the Israel-Darmois matching conditions, \cite{Eisenhart}, a homogeneous and isotropic cosmological metric with a de Sitter - Schwarzschild like metric can be successfully matched across a spherical 3-surface $\Sigma$ which is at fixed coordinate radius of the cosmology metric frame (see also \cite{Balbinot:1988zc}).  However, the radius of the surface in the Black Hole (BH) frame  is time evolving.  
The matching requires the first fundamental form, intrinsic metric, and second fundamental form
		(extrinsic curvature), calculated in terms of the coordinates on $\Sigma$, to be equal (opposite for the second fundamental form)
		on both sides of the surface \cite{matching}, \cite{Dyer:2000sn}.
		
		A homogeneous isotropic cosmological metric can be written in spherical coordinates as
		\begin{equation}
			ds^2=-dt^2+a^2(t)\left[r^2\left(d\theta^2+\sin^2\theta \,  d\phi^2\right)                       +(1-\kappa \,r^2)^{-1}\,dr^2\right]\, ,
			\label{eq:FRW}
		\end{equation}
		where $a(t)$ is the scale factor and $\kappa=0, \pm 1$ is the spatial curvature constant.
		
		The junction conditions \cite{Eisenhart} allow us to use different coordinate systems on both sides of the
		hypersurface. Thus, the cosmology exterior metric (\ref{eq:FRW}) can be joined smoothly
		to the following quantum-modified Schwarzschild metric,\cite{Koch:2014cqa},
		\begin{equation}
			ds^2=-F(R)\,dT^2
			+R^2\left(d\theta^2+\sin^2\theta d\phi^2\right)
			+F(R)^{-1}\,dR^2\, 
			\label{generalBH}
		\end{equation}
		where F(R) is defined by 
			\begin{equation}
			F\left( R \right) = 1 - \frac{2\,G_{k}\,M}{R} - \frac{\Lambda_{k}\,R^{2}}{3}.
			\label{BHDS}
		\end{equation}
		Note, that now the Newton constant and the cosmological constants are not constants but functions of one characteristic length scale  
        $\sim 1/k$. Both will be functions of $R$ as we shall explain later.
				The first fundamental form is the metric on $\Sigma$ induced by the
		spacetime in which it is embedded. This may be written as
		\begin{equation}
			\gamma_{\alpha\beta}=g_{ij}\frac{\p}{\p u^\alpha}x^i
			\frac{\p }{\p u^\beta}x^i\, ,
		\end{equation}
		where $u^\alpha=(u^1\equiv u,\, u^2\equiv v,\, u^3\equiv w)$ 
		is the coordinate system
		on the hypersurface. Greek indices run over $1,\ldots ,3,$ while Latin indices over
		$1,\ldots ,4.$\\
		The second fundamental form \cite{Eisenhart} is defined by
		\begin{equation}
			K_{\alpha\beta}=n_{i;j}\,\frac{\p }{\p u^\alpha}x^i\frac{\p }{\p u^\beta}x^i=(\Gamma^p{}_{ij}n_p-n_{i,j})
			\frac{\gamma_{\alpha\beta}}{g_{ij}}\, ,
			\label{eq:SFF}
		\end{equation}
		where $n_a$ is a unit normal to $\Sigma$ and $\Gamma^p{}_{ij}$ are the Christoffel symbols. We use subscripts $F$ and $S$ to denote quantities associated with the FLRW and
		modified Schwarzschild metrics, respectively \cite{Dyer:2000sn}.
		
Let us consider  a spherical hypersurface $\Sigma$ given by the function
$f_F(x_F^i)=r-r_\Sigma=0,$ where $r_\Sigma$ is a constant.
This hypersurface in the FLRW frame can be parametrized by
$x_F^i\,=\, ( t=u,\, \theta=v,\, \phi=w,\,r=r_\Sigma )$, while
the parametrization in the BH frame is
$x_S^i=\left ( T=T_S(u),\,\theta=v,\, \phi=w,\,R=R_S(u)\right )$. The fact that we model $R=R_S(u)$ implies that the hypersurface $\Sigma$ may not remain in constant BH radial distance as the universe expands. This radial distance is also called \sch  radius.
The successful matching will prove that this choice of the matching surface was the appropriate thing to do for successful modeling of the junction. \emph{From now on wherever in this section we write $T$ and $R$ we mean $T_S$ and $R_S$ respectively}. 
The first Darmois condition $\gamma_{F\alpha\beta}=\gamma_{S\alpha\beta}$ gives
\begin{equation}
	-1=-F(R)\,\left(\frac{dT}{du}\right)^2
	+F(R)^{-1}\left(\frac{dR}{du}\right)^2\, ,
	\label{eq:First FF1}
\end{equation}
and
\begin{equation}
	a^2\,r_{\Sigma}^2=R^2\, .       \label{eq:First FF2}
\end{equation}
where as we have explained, $R=R_S$. The last equation  \ref{eq:First FF2} is an important one and describes that the matching takes place in the hypersurface $\Sigma$ where the BH radial distance is the time dependent \sch radius $R_S$ and the cosmological coordinate distance is the constant $r=r_{\Sigma}$.

The second Darmois condition requires to estimate the two second fundamental forms. First the normal vector ,$n_i$, to the spherical hypersurface $\Sigma$
has to be evaluated.
If $\Sigma$ is given by the function $f[x^a(u^\alpha)]=0$, then $n_i$ can be
calculated from
\begin{equation}
	n_i=-\frac{f_{,i}}{\sqrt{|g^{ab}f_{,a}f_{,b}|}}\, , \label{eq:unit norm eq}
\end{equation}
where $,i$ denotes $\frac{\p}{\p x^i}$\,.

The (outward pointing) unit normal in the FLRW frame can be calculated from Eq. (\ref{eq:unit norm eq})
and $f_F(x_F^i)=r-r_{\Sigma}=0.$ The result is
\begin{equation}
	n_{Fi}=(0,0,0,-(\frac{a^2}{1-k\,r_{\Sigma}})^{1/2} ).
	\label{nfi}
\end{equation}
Note, that the unit normal is spacelike, i.e. $n_F^i n_{Fi}=+1$.
Since the equation of the matching surface $f_F(x_S^i)$ is uknown in the Schwarzschild frame the normal vector $n_S^i$ cannot be calculated directly from
Eq. (\ref{eq:unit norm eq}).
A condition for $n_{S}^{i}$ can be derived from the partial differentiation
of $x_S^i(u^\alpha)$ with respect to $u^\alpha$ which generates a tangential vector to the hypersurface $\Sigma$.
\begin{equation}
	n_{Si}\frac{\partial x_S^i}{\partial u^\alpha}=0\, ,
	\label{eq:second cond}
\end{equation}
From (\ref{eq:second cond}) one obtains  $n_{S2}=n_{S3}=0$ and
\begin{equation}
	n_{S1}\frac{dT}{du}+n_{S4}\frac{dR}{du}=0.
	\label{S2cond}
\end{equation}
Furthermore, $n_{S}^{i}$ must satisfy the identity
\begin{equation}
	n_S^i n_{Si}\equiv n _F^i n_{Fi}=1\;\;
	\label{eq:first cond}
\end{equation}
Then (\ref{eq:first cond}) gives
\begin{equation}
	-\left(F(R)\right)^{-1}{n_{S1}}^2 +F(R){n_{S4}}^2=1.
	\label{eq:normal S1 cond}
\end{equation}

Now using equation (\ref{eq:First FF1}) and equations (\ref{eq:normal S1 cond}), (\ref{S2cond}) it is possible
to evaluate $n_{Si}$ as a function of $u^\alpha:$
\begin{equation}
	n_{Si}=\left(\mp\frac{dR}{du}, \; 0, \; 0, \;
	\pm\frac{dT}{du}\right)\, .   \label{eq:normal K}
\end{equation}

The second fundamental form can be easily calculated in the FLRW frame due to the simple form of $n_{Fi},$ Eq. (\ref{nfi}). Thus
from (\ref{eq:SFF}) we get
\begin{eqnarray}
	K_{F\alpha\beta}
	& = & \Gamma^4_{Fij}n_{F4}
	\frac{\p x^i_F}{\p u^\alpha}\frac{\p x^j_F}{\p u^\beta}
	-n_{F4,j}\frac{\p x^4_F}{\p u^\alpha}\frac{\p x^j_F}{\p u^\beta}
\end{eqnarray}
However, $ \frac{\p x^4_F}{\p u^\alpha}=\frac{\p r_\Sigma}{\p u^\alpha}=0 $, and
\begin{equation}\label{qwsa}
	\frac{\p x^i_F}{\p u^\alpha}\frac{\p x^j_F}{\p u^\beta}=\frac{\p x^\mu_F}{\p u^\alpha}\frac{\p x^\nu_F}{\p u^\beta}=\delta^\mu_\alpha\delta^\nu_\beta
\end{equation}
so the second fundamental form becomes
\begin{eqnarray}
	K_{F\alpha\beta}
	& = & n_{F4}\Gamma^4_{F\alpha\beta} \nonumber \\
	& = & -\frac{1}{2}|g_{F44}|^{1/2}g_F^{4i}
	(g_{F\alpha i,\beta}+g_{F\beta i,\alpha}-g_{F\alpha\beta,i})\nonumber \\
	& = & \frac{1}{2}|g_{F44}|^{1/2}g_F^{44}g_{F\alpha\beta,4}\, ,
\end{eqnarray}
and finally
\begin{equation}
	K_{F\alpha\beta}=\frac{1}{2}\left(\frac{a^2}{1-\kappa\,r^2}\right)^{-{1/2}} g_{F\alpha\beta,4}\, .\label{eq:Simple SFF}
\end{equation}
From equation (\ref{eq:Simple SFF}) it is obvious that the only non zero components of $K_{F\alpha\beta}$
are $K_{F 2 2}$ and $K_{F 3 3}$

The estimation of the second fundamental form of $\Sigma$ in the Schwarzschild frame is more complicated.
First we have to calculate $n_{Si}$ as a function of $x_S^i.$ If we once more differentiate Eq. (\ref{eq:second cond}) with
respect to $u^\alpha$ it follows that
\begin{equation}
	n_{Si,j}\frac{\p x_S^i}{\p u^\alpha}\frac{\p x_S^j}{\p u^\beta}=
	-n_{Si}\frac{\p ^2 x_S^i}{\p u^\alpha\p u^\beta}\, ,
\end{equation}
which with the help of the definition equation \ref{eq:SFF}, gives
\begin{equation}
	K_{S\alpha\beta}=\Gamma^p_{Sij}n_{Sp}
	\frac{\p x_S^i}{\p u^\alpha}\frac{\p x_S^j}{\p u^\beta}
	+n_{Si}\frac{\p ^2 x_S^i}{\p u^\alpha\p u^\beta}\, .
	\label{eq:SFF new}
\end{equation}
Using equation (\ref{eq:SFF new}), we find that
$K_{S\alpha\beta}=0 ,\,\forall\,\alpha\neq\beta.$ This is true since the second derivative is zero for $\alpha\neq\beta$ and the first term of equation (\ref{eq:SFF new}) also is zero for $\alpha\neq\beta$ due to the specific Christoffel symbols.

The second Darmois matching condition $K_{S\alpha\beta}=K_{F\alpha\beta}$ on $\Sigma,$
provides the following differential equations
\begin{eqnarray}
	K_{S11} &\equiv & \Gamma^4_{S11}n_{S4}\left(\frac{dT}{du}\right)^2
	+\Gamma^4_{S44}n_{S4}\left(\frac{dR}{du}\right)^2
	+2\Gamma^1_{S14}n_{S1}\frac{dT}{du}\frac{dR}{du}
	\nonumber \\
	&       &  \ +n_{S1}\frac{d^2T}{du^2}+n_{S4}\frac{d^2R}{du^2}
	=0\equiv K_{F11}\, , \label{eq:SFF1}
\end{eqnarray}
since $p$ can be either 1 or 4 for non zero $n_{Sp}$ and the only relevant non zero Christofell symbols $\Gamma^1_{Sij}$, $\Gamma^4_{Sij}$  are only $\Gamma^1_{S14},\,\Gamma^1_{S41}$ and all $\Gamma^4_{Sii}$. It is also
\begin{equation}
	K_{S22} = K_{F22} \Leftrightarrow \Gamma^4_{S22}n_{S4}=a\,r_\Sigma \sqrt{1-\kappa\,r_\Sigma^2}   \label{eq:SFF2}
\end{equation}
and
\begin{equation}
	K_{S33}= K_{F33} \Leftrightarrow  \Gamma^4_{S33}n_{S4}=a\,r_\Sigma\sin^2\theta \,\sqrt{1-\kappa\,r_\Sigma^2}\, .   \label{eq:SFF3}
\end{equation}
Now the relevant Christoffel symbols are $
\Gamma^1_{S14} =  \frac{F'}{2 F},\, \,
\Gamma^4_{S44}  =  -\frac{F'}{2 F},\, \,
\Gamma^4_{S11}  =  \frac{1}{2} F \left(F'\right),\,\,
\Gamma^4_{S22}  =  -R\, F,\, \,
\Gamma^4_{S33}  =  -\sin^2\theta\,R\,F,\, $
where notation "$'$" is understood as derivative with respect to $R$.

Equations (\ref{eq:SFF2}) and (\ref{eq:SFF3}) being equivalent, we can use
either of them and Eq. (\ref{eq:First FF2}) to obtain
\begin{equation}
	\frac{dT}{du}=\mp\frac{\sqrt{1-\kappa\,r_\Sigma^2}}{1 - \frac{2\,G_{k}\,M}{R} - \frac{\Lambda_{k}\,R^{2}}{3}}\, .
	\label{eq:time der}
\end{equation}
It then follows from Eq. (\ref{eq:First FF1}) that
\begin{equation}
	\left(\frac{dR}{du}\right)^2 =\frac{2\,G_{k}\,M}{R} + \frac{\Lambda_{k}\,R^{2}}{3}-\kappa\,r_\Sigma^2\, .\label{eq:rho der}
\end{equation}
%\begin{equation}
	%\left(\frac{dR}{du}\right)^2 = (1-k\,r_\Sigma^2)-1+\frac{2M(R)}{R}\, %.\label{eq:rho der}
	%\end{equation}
Now differentiating eqs. (\ref{eq:time der}) and (\ref{eq:rho der}) w.r.t. to $u$, we get the following expressions
\begin{equation}
	\frac{d^2T}{du^2}=\pm \sqrt{1-\kappa\,r_\Sigma^2}\, \frac{d\Big(+\frac{2G_k M}{R}+\frac{1}{3}\Lambda_k R^2\Big)}{dR}\, \Big(1-\frac{2G_k M}{R}-\frac{1}{3}\Lambda_k R^2\Big)^{-2}  \, \frac{dR}{du} \,.
\end{equation}
and
\begin{equation}
	\frac{d^2R}{du^2}=\frac{1}{2} \,\frac{d\Big(\frac{2G_k M}{R}+\frac{1}{3}\Lambda_k R^2\Big)}{dR}
	\label{eq:rho 2der}
\end{equation}

%\begin{equation}
	%\frac{d^2T}{du^2}=\pm\frac{\sqrt{1-k\,r_\Sigma^2}}{F(R)^2}\frac{dF}{du}\,.
	%\end{equation}
%and
%\begin{equation}
	%\frac{d^2R}{du^2}= \frac{M'(R)\,R-M(R)}{R^2} . 
	%\label{eq:rho 2der}
	%\end{equation}

With equations (\ref{eq:time der})-(\ref{eq:rho 2der}), Eq. (\ref{eq:SFF1}) is
now identically satisfied. Thus, both the first and second fundamental forms
are continuous on $\Sigma$.

Note that the cosmic evolution through the swiss-cheese is determined by the cosmic evolution of the matching surface that is happening at spherical radius, in the black hole frame, $R_{S}$ (and constant $r_\Sigma$ for cosmic frame). This quantity  enters the differential equations. 

Finally, the matching requirements provide the following equations that are relevant for the cosmological evolution. \emph{From this point and to the rest of the paper we use again separately the symbol $R$ for the BH radial distance and the symbol $R_{S}$ for the Shucking radius},
\begin{eqnarray}
			R_{S}&=&ar_{\Sigma}\label{M1}\\
			\Big(\frac{dR_{S}}{dt}\Big)^{2}&=&1\!-\!\kappa r_{\Sigma}^{2}-\Big(1-\frac{2G_k(R_S)\, M}{R_S}-\frac{1}{3}\Lambda_k(R_S)\, R_S^2\Big)
			\label{M2}\\
			2\frac{d^{2}R_{S}}{dt^{2}}&=&\frac{d\Big(\frac{2G_k(R) \,M}{R}+\frac{1}{3}\Lambda_k(R)\, R^2\Big)}{dR}|_{R_{S}}
			\label{M3}
\end{eqnarray}

		The matching radius $r_{\Sigma}$ can be also understood as the boundary of the volume of the interior solid
		with energy density equal to the cosmic matter density $\rho$. Thus, the interior energy content should equal the
		mass $M$ of the astrophysical object, i.e.
		\begin{equation}
			r_{\Sigma}=\frac{1}{a_0}\Big(\frac{2G_{\!N}M}{\Omega_{m0}H_{0}^{2}}\Big)^{\!\frac{1}{3}}\,,
			\label{lews}
		\end{equation}
		with density parameter $\Omega_{m}=\frac{8\pi G_{\!N}\rho}{3H^{2}}=
		\frac{2G_{\!N}M}{r_{\Sigma}^{3}a^{3}H^{2}}$ and $a_{0}$ set to be 1 for the rest of the analysis.
		
		A rather important information in the context of the aforementioned approach is the exact physical element of our model, galaxies, clusters of galaxies or even super-clusters. In order to answer this, we should look at the physical interpretation of the \sch radius, $r_{\Sigma}$, and its relation with measurable scales of the above astrophysical objects. For Milky Way ($M_{MW} \sim 4 \cdot 10^{11} \ M_{\odot}$) with typical values of cosmological parameters ($\Omega_{m0} \sim 0.3, \ H_0 \sim 67$ km/Mpc/s, \cite{Aghanim}) using Eq.(\eqref{lews}) we have $ r_{\Sigma} \sim 1.4$ Mpc where the astrophysical radius is $R_{as} \sim 0.15$ Mpc. Similarly, for the Local Group $M_{LG} \sim (7-20) \cdot 10^{11}$ we find $r_{\Sigma} \sim 2.34$Mpc and the astrophysical radius is $R_{as} \sim 1.53$ Mpc. Lastly, for the case of our local super-cluster, VIRGO ($M_{VSC} \sim 10^{15}\ M_{\odot}$) it is $R_{as} \sim 30.67$ Mpc and $	r_{\Sigma} \sim 4$ Mpc. We observe that only in the scale of galaxy clusters it is $	r_{\Sigma}/R_{as}$ of order $\mathcal{O}(1)$, thus thereafter we will consider galaxy clusters as the elementary entities of the model. 
		
		Eqs. (\ref{M1}, \ref{M2}, \ref{M3}) for constant $G_{k}$ and $\Lambda_{k}$ reduce to the conventional Friedman-Lemaitre-Robertson-Walker (FLRW) expansion equations. However, although these equations generate the standard cosmological equations of the
		$\Lambda\text{CDM}$ model, even in this simple case the interpretation is different since the $\Lambda$ term appearing in the black hole metric,
		\cite{Koch:2014cqa}, is like an average of all anti-gravity sources inside the \sch radius of a galaxy or cluster of galaxies. These anti-gravity sources one can claim that may arise inside astrophysical black holes in the centers of which the presence of a quantum repulsive pressure could balance the attraction of gravity to avoid the not desired singularity, or they arise due to IR quantum corrections of a concrete quantum gravity theory. Furthermore, even for constant $G_{k}$ and $\Lambda_{k}$ the coincidence problem can be removed if the anti-gravity sources, due to the local effective positive cosmological constant, are related with the large scale structure that appears recently. In our analysis, we allow the running of $G_{k}$ and $\Lambda_{k}$ which is the most natural thing if there are IR quantum gravity corrections.

\section{IR fixed point and cosmic acceleration}
\label{III}
%For studying the cosmological evolution and its characteristics we recall that in section II the infrared AS corrections to the Newton and cosmological constants result to eqs. (\ref{ir1}) which we are going to use them with a more short notation as $G_k={G(k)}_{{IR}} = \frac{g_{*}}{k^{2}} + h_{1}k^{\theta_{1} - 2}$ and  $\Lambda_k=\Lambda\left( k \right)_{{IR}} = \ \lambda_{*}k^{2} + h_{2}k^{\theta_{2}+2}$ .
In this section the basic idea of the IR fixed point RG scaling will be introduced and 
the cosmic evolution of this model will be discussed.
\subsection{The IR fixed point scaling}
The presence of an IR instability in the RG flow for $G$ and $\Lambda$ is deeply rooted 
in the structure of the fluctuations determinant associated to the Einstein-Hilbert
action
\begin{equation}
    S_k=\frac{1}{16 \pi G_k}\int dx^4 \sqrt{g} (-\mathcal{R} + 2\Lambda_k).
\end{equation}
Let us consider a family of off-shell spherically symmetric contributions
labelled by the radius $\phi$ in order to distinguish the contributions from
the two operators $\sqrt{g}\propto \phi^4$ and $\sqrt{g} \mathcal{R}\propto \phi^2$
to the Einstein-Hilbert flow \cite{Reuter:2004nx}. It is possible to decompose the fluctuating 
metric $h_{\mu\nu}$ to the sphere into its irreducible pieces in terms of the 
corresponding spherical harmonics. In particular, in the transverse-traceless 
sector the operator $S^{(2)}_k$ is given by  
\begin{equation}
    -g^{\mu\nu}\nabla_\mu\nabla_nu + 8 \phi^{-2}+k^2- 2\Lambda_k
\end{equation}
up to a positive constant. If the 
spectrum of the covariant laplacian $-g^{\mu\nu}\nabla_\mu\nabla_nu$ is non-negative, 
a singularity develops at a finite $k$ in the case of a {\it positive} cosmological
constant. 
  
In a scalar field theory an identical mechanism is present when the hessian of the action
vanishes in the broken phase. In this case perturbation theory fails and the path-integral
is dominated by non-homogeneous configurations \cite{Ringwald:1989dz} that produces
a ``tree-level" instability  renormalization already at level of the bare action \cite{Alexandre:1998ts}. 
In spite of the complexity of the vacuum structure, the approach based 
on functional flow equation is able to sum an infinite tower of (irrelevant UV) operator
and it reproduces the correct flattening of the effective potential $U(\phi)$ in the broken phase
\cite{Bonanno:2004pq}. In this case, below the coexistence line, the potential behaves
as $U(\phi)\propto - k^2 \phi^2$ and approaches a flat bottom as $k\rightarrow 0$
\cite{Berges:2000ew}.

In gravity the situation is more difficult because of the technical difficulties in solving the RG 
flow equation beyond the Einstein-Hilbert truncation in the $k\rightarrow 0$ limit
\cite{Benedetti:2012dx,Demmel:2014sga,Dietz:2012ic}.
%CITE works with f(R) theory 
Based on the 
analogy with the scalar theory one can imagine that a RG trajectory which 
realizes a tree-level renormalization exists \cite{Ohta:2021bkc}.
In particular one can write that in the $k\rightarrow 0$ limit 
the dimensionless Newton constant and cosmological constant
are attracted toward an IR fixed point $(g_{*}, \lambda_{*}$) according to 
the following trajectory
\begin{equation} 
g\left( k \right) = g_{*} + h_{1}k^{\theta_{1}}, \quad \quad
\lambda\left( k \right) = \ \lambda_{*} + h_{2}k^{\theta_{2}}, 
\end{equation}
where $(\theta_{1},\theta_{2} \geq 0)$ are two unknown critical exponents. 
In terms of dimensionful quantities, we thus have
\begin{equation}\label{ir1}
{G(k)}_{{IR}} = \frac{g_{*}}{k^{2}} + h_{1}k^{\theta_{1} - 2} \quad\quad
\Lambda\left( k \right)_{{IR}} = \ \lambda_{*}k^{2} + h_{2}k^{\theta_{2}+2} 
\end{equation}
Here, $k$ should be considered not as a momentum flowing into a loop, but as an inverse of a characteristic 
distance, of the physical system under consideration, over which the averaging of the field variables is performed. 
It can then  be associated to the radial proper distance or the matter density. F

For the rest of the paper in order to study the late cosmic evolution, we assume that the values of Newton constant and cosmological constant at the astrophysical scales will be those suggested in Eq. (\ref{ir1}) i.e. $G_k={G(k)}_{{IR}}$ and  $\Lambda_k=\Lambda\left( k \right)_{{IR}}$ .

In the first case we need to calculate the radial proper distance and in particular the \sch proper radial distance, $D_p(R_S)$, which can be found from
		\begin{equation}
			D_{p}\left( R_S \right) = \ \int_{R_{min}}^{R_{S}}\frac{R}{\sqrt{F(R)}}
		\end{equation}
where $R_{min}$ is zero in our case or a very small value if AS provides a non singular black hole center, see \cite{zarikasBH}. Then, the derivative with respect to $R$ is 
		\begin{equation}
			D_{p}^{'}\left( R_S \right) = \ \frac{1}{\sqrt{F(R_{S})}}
		\end{equation}
		
		Next, from matching equations equations (\ref{M1},\ref{M2},\ref{M3}) we get
		\begin{equation}
			{\left(\frac{dR_{S}}{dt}\right)}^{2} = 1 - \kappa\,r_{\Sigma}^{2} - F(R_{S})
		\end{equation}
which gives
		\begin{equation}
			r_{\Sigma}^{2}{\dot{a}}^{2} = 1 - \kappa\,r_{\Sigma}^{2} - 1 + \frac{2\,G_{k}(R_{S})\,M}{R_{S}} + \frac{1}{3}\Lambda_{k}(R_{S})\,R_S^{2} 
		\end{equation}	
and		
		\begin{equation}\label{cosmo}
			\frac{{\dot{a}}^{2}}{a^{2}} = -\frac{\kappa}{a^{2}}+   \frac{2G_k(R_{S})M}{a^3\,r_{\Sigma}^{3}} + \frac{1}{3}\Lambda(R_{S})
		\end{equation}

\subsection{Scaling using the Cluster Radial proper distance }
Here the scaling we assume uses the proper distance
\begin{equation}
	k =  \frac{\xi}{D_p(z)}
	\label{scale1}
\end{equation}
Using Eq.(\ref{cosmo}) we get
		\begin{equation}
			\frac{{\dot{a}}^{2}}{a^{2}} + k\frac{1}{a^{2}} = \chi + \psi
		\end{equation}
		where
		\begin{equation}
			\chi = 2\left\lbrack \frac{g_{*}}{\xi^{2}}D_{p}^{2} + h_{1}\xi^{\theta_{1} - 2}D_{p}^{2 - \theta_{1}} \right\rbrack M\frac{1}{r_{\Sigma}^{3}a^{3}} 
		\end{equation}
		and
		\begin{equation}
			\psi =  \frac{1}{3}\left\lbrack \lambda_{*}\xi^{2}D_{p}^{- 2} + h_{2}\xi^{\theta_{2} + 2}D_{p}^{- 2 - \theta_{2}} \right\rbrack.
		\end{equation}
		
		One useful quantity for evaluating the dark energy is the rate of change of the proper Shucking distance $D_{p}$. We have 
		$\frac{dD_{p}}{\text{dR}} = \frac{dD_{p}}{\text{dt}}\frac{\text{dt}}{\text{dR}} = \frac{1}{\sqrt{F(R_{S})}}$
		and thus, 
		\begin{equation}\frac{dD_{p}}{dz} = \frac{- r_{\Sigma}}{{(1 + z)}^{2}}\frac{1}{\sqrt{F}}\end{equation}
		where $z$ is the redshift $z=-1+\frac{1}{a}$ while the time derivative is
		\begin{equation}
			{\dot{D}}_{p}  = \frac{r_{\Sigma}\,a\,H}{\sqrt{1 - \frac{\text{2M}}{a\,r_{\Sigma}}\left( \frac{g_{*}}{\xi^{2}}D_{p}^{2} + h_{1}\xi^{\theta_{1} - 2}D_{p}^{2 - \theta_{1}} \right) - \frac{1}{3}\left( a\,r_{\Sigma} \right)^{2}(\lambda_{*}\,\xi^{2}D_{p}^{- 2} + h_{2}\xi^{2 + \theta_{2}}D_{p}^{- 2 - \theta_{2}})}}.
		\end{equation}

		The dark energy and dark matter part into the expansion equations is  
		\begin{equation}
			H^{2} + \frac{k}{a^{2}} =  \frac{8\,\pi\,G_{N}\, \rho_{DM}}{3}  + \frac{8\,\pi\,G_{N}\, \rho_{DE}}{3},
		\end{equation}
		where the dark energy and dark matter density are given by 
		\begin{equation}
			\rho_{DE} =\frac{3}{8\pi\,G_{N}} \left[-\frac{2G_N\,M}{a^3\,r_{\Sigma}^{3}}+\psi+\chi\right], \,\,\,\,\rho_{DM} = \frac{3}{8\pi\,G_{N}}\frac{2G_NM}{a^3\,r_{\Sigma}^{3}}
		\end{equation}.
		The cosmological definition of the dust matter refers to a quality that evolves as $\sim a^{-3}$. In order to use the standard cosmological parameter $\Omega_{m0}$ so to easily compare our models with the concordance one, we extract from the "dark bulk" a portion that evolves as matter. Thus, the remaining one is correctly labeled as Dark Energy.
		Following the common parametrization the expansion can be written as 
		\begin{equation}\label{expansiomega}
			E^2=\frac{H^{2}}{H_0^{2}} =\Omega_{DM} + \Omega_{DE} +\Omega_{\kappa} 
		\end{equation}
		where
		\begin{equation}
			\Omega_{DM} = \frac{\rho_{DM}}{\rho_{cr}} \,\,\
			\text{and}\,\,\,\
			\Omega_{DE} =\frac{\rho_{DE}}{\rho_{cr}}\,\,
			\text{and}\,\,\,\ 
			\Omega_{k} = \frac{-\kappa}{a^2H_0^{2}}
		\end{equation}
		with $\rho_{cr}=\frac{3H_0^2}{8\pi\,GN}$.
		
		The cosmic acceleration, using Eq.(\ref{M3}), is then expressed as follows 
		\begin{equation}
			2\frac{\ddot{a}}{a} + \frac{{\dot{a}}^{2}}{a^{2}} + \frac{\kappa}{a^{2}} = - 8\pi\,G_{N}\,P
		\end{equation}
		where the total pressure is the sum of cosmic fluid pressure plus dark energy pressure  $P=p+p_{DE}$.
		In the case of dust Universe $p=0$, which is valid for small redshifts and  since $H^2 + \kappa/a^2 =\chi+\psi $ we have
		
		\begin{equation}
			2\frac{\ddot{a}}{a} + \chi+\psi= - 8\pi\,G_{N}\,p_{DE}
		\end{equation} 
		Finally, the dark energy coefficient of equation or state is given by \begin{equation}
			w_{DE}= \frac{p_{DE}}{\rho_{DE}}
		\end{equation}

		In order to calculate the expansion of the Universe and the time evolution of the $\Omega_{DM},\, \Omega_{DE},$ the equation of state parameter $w_{DE}$ and the deceleration parameter $q=-\frac{a\ddot{a}}{\dot{a}^2}$ we must solve numerically ODEs for the scale factor and the proper distance. The relavant quantities as functions of the redshift are give n below.

		The derivative with respect to the redshift of the proper distance is 
		\begin{equation}
			\frac{D_p}{dz}=-\frac{r_{\Sigma}}{(z+1)^2 \sqrt{-\frac{2 M (z+1) \left(\frac{g_{*} D_p(z)^2}{\xi ^2}+h_1 \left(\frac{\xi }{D_p(z)}\right)^{\theta_1-2}\right)}{r_{\Sigma}}-\frac{r_{\Sigma}^2 \left(h_2 \left(\frac{\xi }{D_p(z)}\right)^{\theta_2+2}+\frac{\lambda_{*} \xi ^2}{D_p(z)^2}\right)}{3 (z+1)^2}+1}}
		\end{equation}
		This is one of the differential equations we have to solve together with the expansion rate ODE for the scale factor Eq.(\ref{expansiomega})
		
		\begin{equation}
			\begin{split}
				q(z)=&V^{-1} \biggl[ 6 J M D_p(z)^5 \left(h_1 \left(\frac{\xi}{D_p(z)}\right)^{\theta_1}+g_*\right)+\frac{6 \sqrt{3} M r_{\Sigma} D_p(z)^4 \left(h_1 (\theta_1-2) \left(\frac{\xi}{D_p(z)}\right)^{\theta_1}-2 g_*\right)}{z+1} \nonumber \\
				&-\frac{2 J \xi ^4 r_{\Sigma}^3 D_p (z) \left(h_2\left(\frac{\xi}{D_p(z)}\right)^{\theta_2}+\lambda_*\right)}{(z+1)^3}+\frac{\sqrt{3} \xi ^4 r_{\Sigma}^4 \left(h_2 (\theta_2+2) \left(\frac{\xi}{D_p(z)}\right)^{\theta_2}+2 \lambda_*\right)}{(z+1)^4} \biggr] 
			\end{split}
		\end{equation}
		where
		\begin{equation}
			V=
			{2 J \left(6 M D_p(z)^5 \left(h_1\left(\frac{\xi}{D_p(z)}\right)^{\theta_1}+g_*\right)+\frac{\xi ^4 r_{\Sigma}^3 D_p (z) \left(h_2 \left(\frac{\xi }{D_p(z)}\right)^{\theta_2}+\lambda_*\right)}{(z+1)^3}\right)}
		\end{equation}
		and
		\begin{equation}
			J=\sqrt{-\frac{6 M (z+1) D_p(z)^2 \left(h_1 \left(\frac{\xi }{D_p(z)}\right)^{\theta_1}+g_*\right)}{\xi ^2 r_{\Sigma}}-\frac{\xi ^2 r_{\Sigma}^2 \left(h_2 \left(\frac{\xi }{D_p(z)}\right)^{\theta_2}+\lambda_*\right)}{(z+1)^2 D_p(z)^2}+3}
		\end{equation}
		where the Sucking radius for a cluster of galaxies with mass $M$ is
		
		\begin{equation}
			r_{\Sigma}=\sqrt[3]{\frac{2 G_N M}{H_0^2 \Omega_{DM\,0}}}
		\end{equation}
		the dark energy and dark matter part evolution is :
		
		\begin{equation}
			\Omega_{DE}(z) = \frac{ \Omega_{m0} (z+1)^3 \left(\frac{g_* {Dp} (z)^2}{\xi ^2}+h1 \left(\frac{\xi }{{D_p}(z)}\right)^{\theta_ 1-2}\right)} {G_N} +\frac{{h2} \left(\frac{\xi }{Dp(z)}\right)^{\theta_2+2}+\frac{\lambda_* \xi ^2}{Dp(z)^2}}{3\,H_0^2} 
			-\Omega_{m0} (z+1)^3
		\end{equation}

		\begin{equation}
			\Omega_{DM}(z) = \Omega_{m0} (z+1)^3
		\end{equation}
		
		The evolution of the dark energy coefficient is 
		
		\begin{equation}
			\begin{split}
				w_{DE}(z) = & \tilde{V}^{-1}\,r_{\Sigma} \biggl[ \sqrt{3}\, r_{\Sigma}^3\, \xi^4 
				\big( 2 \lambda_*+h_2 (2+\theta_2) (\frac{\xi}{D_p(z)})^{\theta_2} \big) 
				-3 \tilde{J} r_{\Sigma}^2 (1+z) \xi^4 \big[\lambda_*+h_2 (\frac{\xi}{D_p (z)})^{\theta_2}\big] D_p (z) \nonumber \\
				&-6 \sqrt{3} M (1+z)^3 \big(2 g_{*} -h_1 (-2+\theta_1)\left(\frac{\xi}{D_p(z)}\big)^{\theta_1}\right)D_p (z)^4 \biggr]
			\end{split}
		\end{equation}
		where
		
		\begin{equation}
			\tilde{V} =	3 \tilde{J} (1+z)^4 \biggr[  r_{\Sigma}^3 \xi^4 \big(\lambda_{*} + h_2 \big( \frac{\xi}{D_p (z)} \big) ^{\theta_2} \big) D_p(z)/(1+z)^3 - 6 G_N M \xi^2 D_p (z)^3 + 6 M \big( g_{*} +h_1 \big( \frac{\xi}{D_p(z)}\big)^{\theta_1}\big) D_p(z)^5 \biggl]
		\end{equation}
		
		and
		\begin{equation}
			\tilde{J}= \left\lbrace 3-\frac{ r_{\Sigma}^2 \xi^2 [\lambda_*+h_2 (\frac{\xi}{D_p(z)})^{\theta_2}]}{(1+z)^2 D_p (z)^2}-\frac{6 M (1+z) [g_*+h_1 (\frac{\xi}{D_p(z)})^{\theta_1}] D_p (z)^2}{r_{\Sigma} \xi^2}\right\rbrace ^{1/2}
		\end{equation}

		\subsection{Scaling using the Cluster Density Profile }
		Here, in this subsection we use another scaling,
		\begin{equation}
			k =  \tilde{\xi}\,\rho_{cl}^{1/4}
			\label{scale2}
		\end{equation}
		where the density of the cluster $\rho_{cl}$ is 
		
		\begin{equation}
			\rho_{cl}=\frac{\rho_c}{\left(\frac{\tilde{R}}{\alpha }\right)^{\beta } \left(\frac{\tilde{R}}{\alpha }+1\right)^{3-\beta }}
		\end{equation}
		where $\beta=1/2$ and $\alpha=226\, kpc$, and $\rho_c=5.37\times10^6\,\,M_\odot\,kpc^{-3}$ .
		
		The above density profile contains the standard  Navarro - Frenk - White profile explicitly at $\beta = 1$ and asymptotically at $ \tilde{R}_{core} \rightarrow 0$.
		The parameter $\beta$ describes the central cusp (i.e. $\rho_{cusp} \sim \tilde{R}^{-\beta}$) and it holds $0 < \beta < 3 $,  \cite{Zhao:1995cp}. Following \cite{Newman:2012nw}, where the aforementioned density profile was fitted at observational data, we take their ensemble mean value of $\beta$, that is $\beta=0.5$. The parameter $\alpha$, is the radius of the central cusp and we take $\alpha=226\, kpc$  Lastly, we assume $\rho_c=5.37\times10^6\,\,M_\odot\,kpc^{-3}$. This profile was used as a typical example.

		To proceed further we set $\tilde{R}=D_p$ and using the matching equations we find that the derivative of the proper distance is given now by
		\begin{equation}
			D_p(z)/dz=-\frac{r_{\Sigma}}{(z+1)^2 \sqrt{-\frac{2 M (z+1) A}{r_{\Sigma}}-\frac{r_{\Sigma}^2 B}{3 (z+1)^2}+1}}
		\end{equation}
		
		The $\Omega_{DM}$ remains the same but the $\Omega_{DE}$ is given by
		
		\begin{equation}
			\Omega_{DE}=\frac{2 A M (z+1)^3}{H_0^2 r_{\Sigma}^3}+\frac{B}{3 H_0^2}-\Omega_{m0} (z+1)^3
		\end{equation}
		
		where
		\begin{equation}
			A=\left(\frac{g_{*}}{\tilde{\xi} ^2 \sqrt{\frac{\rho_{c}}{\sqrt{\frac{D_{p}(z)}{\alpha }} \left(\frac{D_{p}(z)}{\alpha }+1\right)^{5/2}}}}+h_{1} \left(\tilde{\xi}  \sqrt[4]{\frac{\rho_{c}}{\sqrt{\frac{D_{p}(z)}{\alpha }} \left(\frac{D_{p}(z)}{\alpha }+1\right)^{5/2}}}\right)^{\theta_{1}-2}\right)
		\end{equation}
		and 
		\begin{equation}
			B=h_{2} \left(\tilde{\xi}  \sqrt[4]{\frac{\rho_{c}}{\sqrt{\frac{D_{p}(z)}{\alpha }} \left(\frac{D_{p}(z)}{\alpha }+1\right)^{5/2}}}\right)^{\theta_{2}+2}+\lambda_{*} \tilde{\xi} ^2 \sqrt{\frac{\rho_{c}}{\sqrt{\frac{D_{p}(z)}{\alpha }} \left(\frac{D_{p}(z)}{\alpha }+1\right)^{5/2}}}
		\end{equation}

		Unfortunately, expressions for $q(z)$ and a $w_{DE}(z)$ are quite long and thus are not presented here for economy of space.

\section{Numerical Results}
In the following, we discuss the ability of our models to reproduce known observable quantities. Specifically, we check if there exist sets of free parameters that are able to reproduce the observed Hubble evolution within the acceptable range variation of the gravitational constant over time and the recent passage to cosmic acceleration. The existence of allowed range of parameters at the parameter space does not guarantees the fitting quality of our models, however the non-existence implies the non-viability of our models. The full assessment of viability for our models ought to be performed using likelihood analysis and model selection methods on cosmological data-sets, such as Supernovae Ia \cite{sn,sn2,sn3,sn4}, Baryonic Accoustic Oscillations \cite{bao1} and direct measurements of the Hubble rate, in the same fashion with \cite{sc3}, will be resented in a forthcoming paper. This analysis is beyond the scope of the current work. 

In what follows, we explicitly assume $\kappa=0$, based upon observational results, i.e CMB analysis \cite{Aghanim}. In any case, even if the Universe is not exactly flat, the contribution of $\Omega_k$ to the expansion is negligible for recent red-shifts like the ones that are relevant in our context.

We have decided to explore the free parameters space working with some simplified but reasonable assumptions. 
One should take into account that the evolution of $G_{k}$, must be smooth enough to comply with observational tests, i.e. solar system constraints, \cite{Fienga:2014bvy, Hofmann:2018myc} and astroseismology, \cite{Bellinger:2019lnl}. A particular way to achieve this is to consider the first term in Eq.(\ref{ir1}) to be subdominant to the second, thus allowing $G(k)$ to remain almost constant and have little dependence of $k$. In the same lines of reasoning, $h_1$ must be very close to $G_N$. Therefore  in our numerical assessments we set $h_1=G_N$  and assume that $\theta_1$ must be close to $2$.
On the other hand, as it has already been proven that a term proportional to $k^2$,  generates successful phenomenology \cite{sc1,sc2,sc3} we also assume that  $\theta_2$ must be close to $0$, and since both terms are practically proportional to $k^2$, we can choose  $\lambda_{*} \ll 1$.
We further simplify our search for a viable cosmological model assuming

\begin{equation}
g_{*} \ll h_{1}k_{IR}^{\theta_{1}} , \quad
\lambda_{*} \ll h_{2}k_{IR}^{\theta_{2}}
\end{equation}
where $k_{IR}$ is the astrophysical scale. 
There are still many combinations of the remaining free parameters  that can provide the correct phenomenology. A particularly simple way to find a subset of the parameter space that gives  reasonable behavior from the phenomenological viewpoint, is to ensure that 
\begin{equation}
\chi(z\simeq0)  \sim \psi(z\simeq0)	,
\end{equation}
which happens if we use as initial condition for the differential equation that provides the evolution of the proper distance $D_P(z)$, the value from the requirement to have as $H(z=0)$ the current measured value of the Hubble rate. In general, the initial value that one gets for the proper distance in order to achieve the current value for $H(0)$ or $\Omega_{DE}$ to be the same as in the concordance model, could possibly give an unnatural value for the proper distance. In contrast, we obtain reasonable values that are of the same order and little bit larger than the \sch radius of real astrophysical objects (clusters of galaxies) which is the correct thing. 
Furthermore, in order not to have new scales or fine tuning we always give values close to order of one O(1) for $\xi$ and the dimensionless $\tilde{\gamma}$ which is related to $h_2$ from the definition $\tilde{\gamma}=\gamma\,G_N^{1-b/2}$. 
%Note also that it is reasonable to   expect $\xi$ of order of one in order from general arguments. It is worth mentioning that the coefficient $\tilde{\xi}$ cannot be of order one, due to the scaling law Eq. (\ref{scale2}).
%In the latter case the length scale that is generated from Eq.(\ref{scale2}) is not the astrophysical cluster length scale that miraculously is the one that provides the correct amount of dark energy today and at the same time miraculously is the length scale that makes the swiss-cheese model an acceptable approximation, since the \sch matching radius is close to the cluster length.
Note, that in general the values of $\xi$ an $\tilde{\xi}$ can be different as Eq.(\ref{scale2}) entails a different scaling length. However, in the case of the first scaling using the proper distance and not the density, the natural length that appears in the calculations is the cluster length scale.\emph{ It worths emphasizing at this point, that using this naturally emerging astrophysical cluster length scale (which means also setting $\xi$ of O(1)), we get the correct amount of dark energy today after a recent passage from deceleration to acceleration and at the same time interestingly, this length scale makes the swiss-cheese model an acceptable approximation, since the \sch matching radius is close to the cluster length.}

%In the following part, we explicitly check particular instances of our models with regard to specific cosmological phenomena.
In order to compare the Hubble rate in the context of AS models with the observations, we have to normalize accordingly the Hubble rate, i.e to demand $H(0) = 100h$, where $h$ is dimensionless parameter and $h \in (0,1)$. From this requirement, we can eliminate $\tilde{\gamma}$ from the Hubble rate demanding the matching to happen in $2r_{\Sigma}$, or we can have an initial condition for the $D_p(0)$ equal a value close to $r_{\Sigma}$ and fix $\tilde {\gamma}$. Thus, we can further manage to lower the number of free parameters for our models, which is beneficial from both conceptual and numerical viewpoints.
In fact, we can also check if a particular instance of the model could describe the accelerated expansion of the universe, while also satisfying the available observational constraints on $\dot{G}/G$. In Fig. (\ref{fig:figure3}) we illustrate the Hubble rate for both scaling relations, with the total of 3 parameters, that are ($\xi, \theta_1, \theta_2$) for the first scaling relation and ($\tilde{\xi}, \theta_1, \theta_2$) for the second scaling relation. 
An indicative set of parameters that can give the desired phenomenology are listed in Table (\ref{T2}).  

We aslo plot the most recent data compilation, that have obtained via the Cosmic Chronometers (CC) method, thus being approximately model-independent, compiled at  \cite{Yu:2017iju}. In addition we depict the Hubble rate of the concordance model (dashed black line) corresponding to the parameters  set $\{\Omega_{m0}, h\} = \{0.3,0.68\}$, to allow for qualitative comparison between the two models. The Hubble rate for each scaling relation is depicted with the solid line. Examination of Fig. (\ref{fig:figure3}) shows that both scenarios can give Hubble rate in close agreement with the observations. 
 
 %We set the initial condition for the numerical solution of Eq. \ref{scale1} to be $2r_s$. The latter is set as such, bearing in mind the relation between Schuking radius and physical distance of the Local Group, discussed previously. Note however that the actual value ought to be determined by fitting the models to observational data, which lies beyond the scope of the current work. %
 
\begin{figure}
\includegraphics[width=0.5\textwidth]{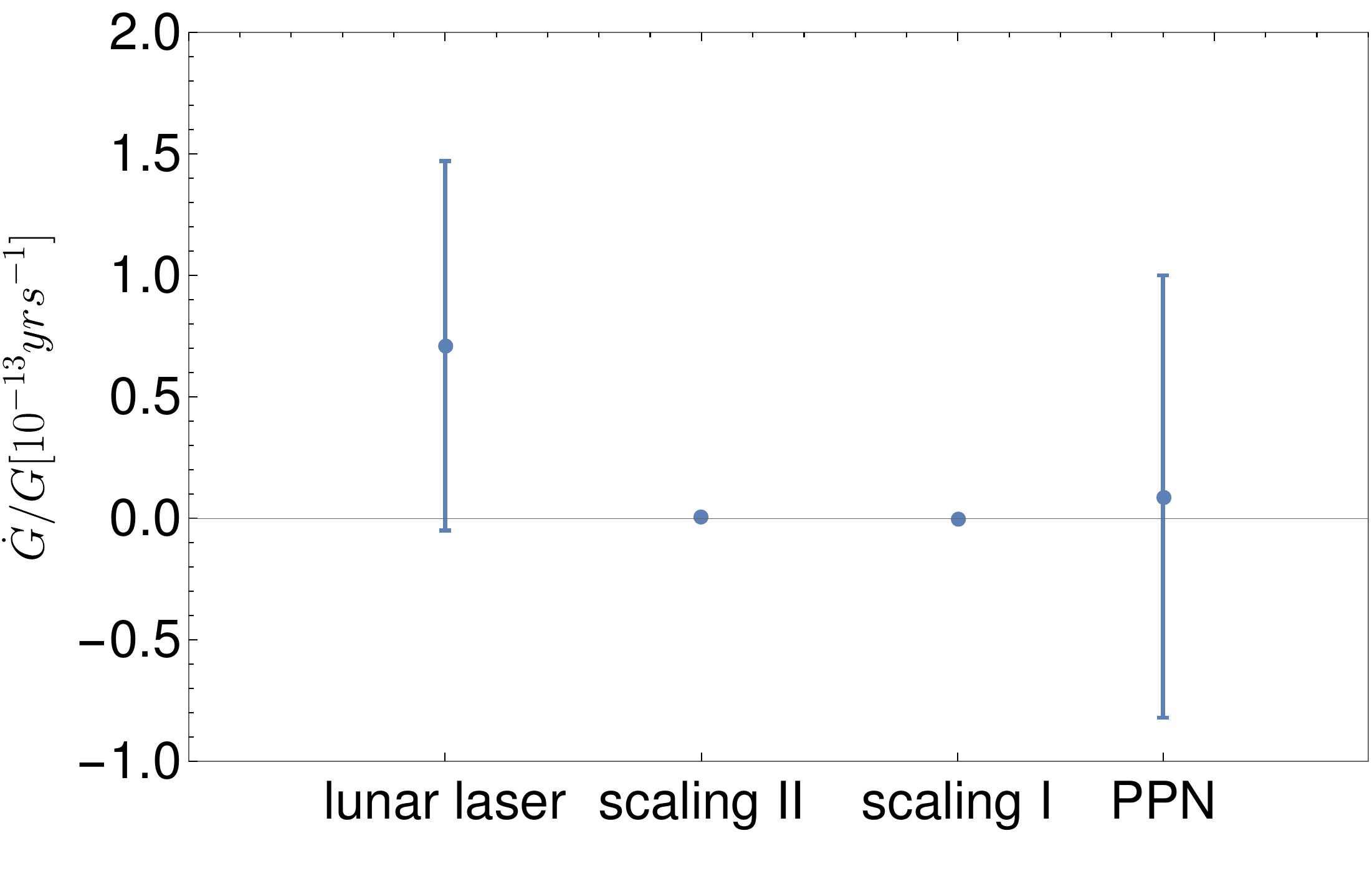}
\caption{Values of the normalized derivative of the Newton constant G around z = 0, for both scaling relations labeled "scaling-I" which is the proper distance scaling and "scaling-II" the scaling using density. We include also the most recent observational values along with their error bars ($1\sigma$). These are: \textbf{(a)} from \cite{Fienga:2014bvy} via calculating possible impact of varying G on planetary orbits (labelled 'solar system orbits') \textbf{(b)} from \cite{Hofmann:2018myc} using lunar laser ranging data (labelled 'lunar laser').}
\label{fig:dot_geff/g}
\end{figure}

 Regarding the value of the normalized derivative of the coupling constant, $\dot{G}/G$ today, we calculate $\dot{G}/G = -0.001 \cdot 10^{-13}yrs^{-1}$ and  $\dot{G}/G = 0.0083\cdot 10^{-13}yrs^{-1}$ for the first and the second scaling relations, respectively. From Fig. (\ref{fig:dot_geff/g}), is easy to observe that both models are within $1\sigma$, that is in full agreement with the observational values from \cite{Fienga:2014bvy, Hofmann:2018myc}. The same with the result from \cite{Bellinger:2019lnl} which is one order of magnitude larger from the other two and is not depicted at Fig. (\ref{fig:dot_geff/g}).
Fig. (\ref{fig:dot_geff/g}) shows the comparison of these results with corresponding observations. As it is apparent from the latter figure, the evolution of G is within 1$\sigma$ of the observed values, for both scaling relations. 

%  , using the parameters .... we obtain ... which lies within 1$\sigma$ of the observations, Fig. (\ref{fig:dot_geff/g}). 

\begin{table}[ht!]
\begin{center}
\begin{tabular}{|| c c c c c||} 
 \hline
  Scaling  Model  & $\xi \,\text{or}\, \tilde{\xi}$ & $\tilde{\gamma}$& $\theta_{1}$ & $\theta_{2}$ \\ [0.5ex] 
 \hline\hline
 Cluster Radial Proper Distance ($\xi, \theta_1, \theta_2$)  &3.0 & 3.0 & 2.001 & 0.06\\
 \hline
$>>$  & 1.0 & 0.1 & 2.005 & 0.05 \\
 \hline
 $\ \ \ $Cluster Density Profile ($ \tilde{\xi}, \theta_1, \theta_2$)  & 1e-28 &10 &2.001 & 0.06 \\ 
 \hline
 $>>$ & 1e-28 & 15 &2.005 & 0.05 \\ 
 \hline
\end{tabular}
\caption{Indicative points on the parameter space for each of the two scaling relations (\S \ref{scale1},\ref{scale2}) that correspond to efficient phenomenology. }
\label{T2}
\end{center}
\end{table}
As another viability test for our models, we use the transition redshift, defined via the requirement $q(z_{tr}) \equiv 0$. The latter corresponds to the moment of the transition between matter era and Dark Energy era in the cosmic history.
In particular, using the plot of the deceleration parameter, q, (second column of Fig.(\ref{fig:figure3}) ) for these particular instances of the two different scaling expressions, we find the transition redshift. For the case of scaling with proper distance, we have $z_{tr}^{I} \simeq 0.8$ which lies within $\sim 3\sigma$ of both \cite{Jesus:2019nnk} results, obtained via a model independent method. For the case of scaling using density, we obtain $z_{tr}^{II} \simeq 1$ which again is statistically compatible with \cite{Jesus:2019nnk} results. From the latter fact we deduce that both models seem to be viable.

\begin{figure}
\centering
\begin{minipage}{.45\linewidth}
  \includegraphics[width=\linewidth]{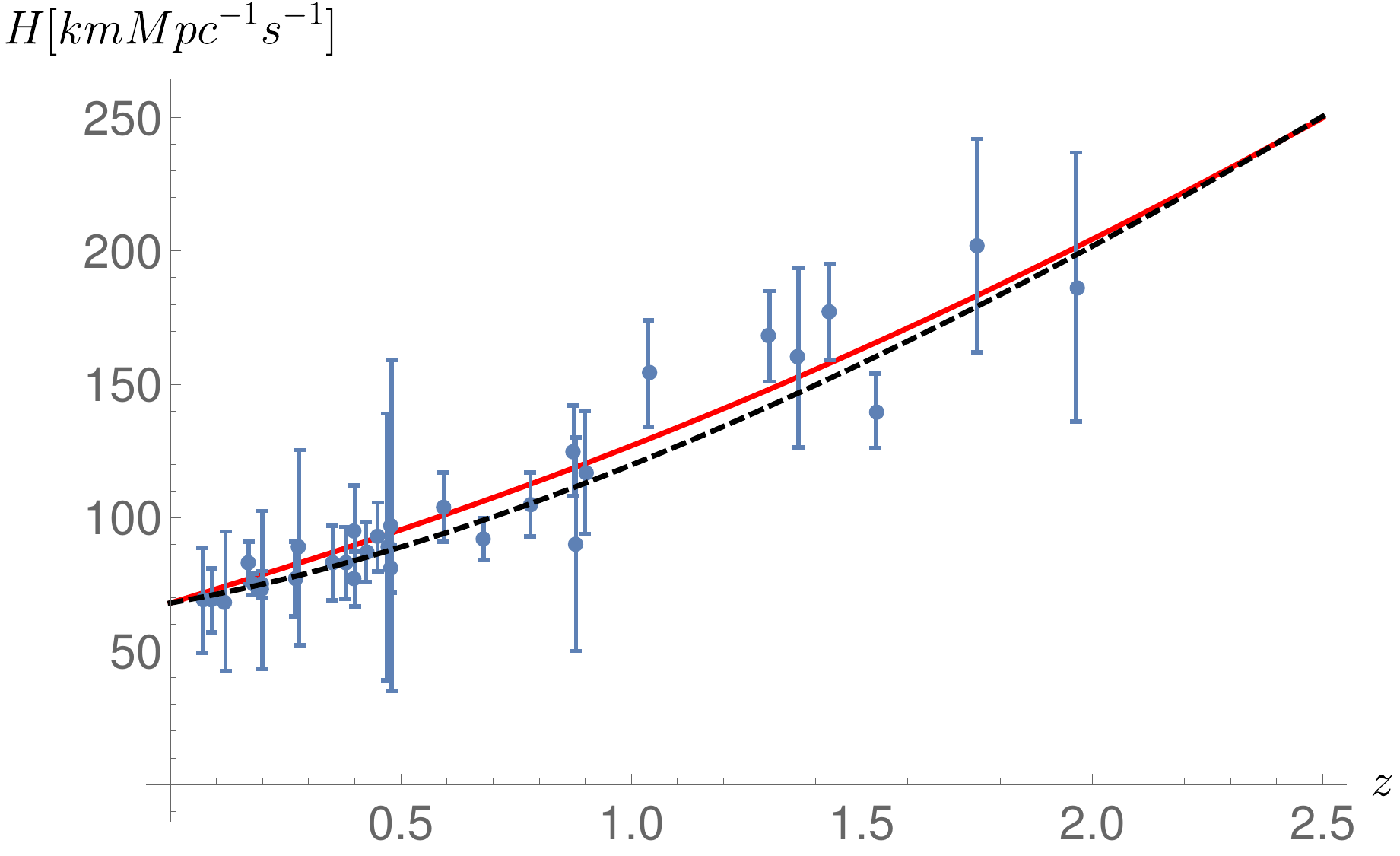}
  %\captionof{figure}{}
  \label{img11}
\end{minipage}
\hspace{.05\linewidth}
\begin{minipage}{.45\linewidth}
  \includegraphics[width=\linewidth]{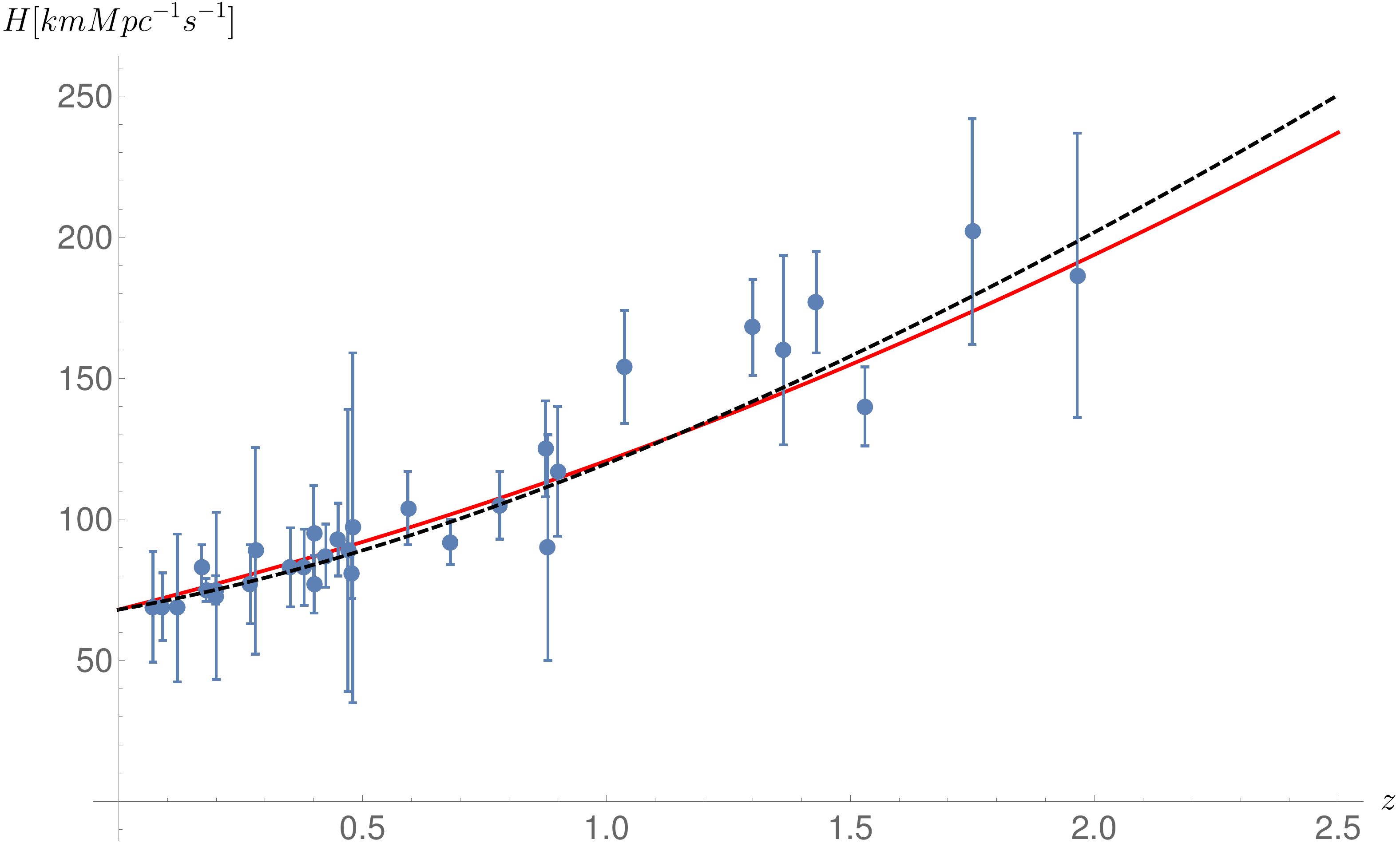}
  %\captionof{figure}{}
  \label{img222}
\end{minipage}
\caption{Plot of the Hubble rate along with a subset of the observational dataset, containing only data points obtained via the Cosmic Chronometers (CC) method, compiled at  \cite{Yu:2017iju}. To allow direct comparison, we depict the Hubble rate of the concordance model (dashed black line) corresponding to the parameters  set $\{\Omega_{m0}, h\} = \{0.3,0.68\}$.  The Hubble rate for each scaling relation is depicted with the solid line. In all cases the initial condition for the solution of proper distance was set to be $2r_s$, for reasons explained in the text. \textbf{Left:} Scaling using proper distance, for the parameter set $\{\Omega_{m0}, \xi, h, \theta_1, \theta_2\} = \{0.3,0.68,1,2.001, 0.06\}$   \textbf{Right:} Scaling using density, for the parameter set $\{\Omega_{m0},h, \tilde{\xi}, \theta_1, \theta_2\} = \{0.3,0.68,10^{-28}, 2.005,0.05\}$
}

\label{fig:scaling-0}
\end{figure}

\iffalse
\begin{figure}[!]
\includegraphics[width=\textwidth]{Plots/f2.png}
\caption{ {{}}
}
\label{fig:scaling-0b}
\end{figure}
%\ayan{Fig. 3 below are TEST PLOTS not FINAL}
\begin{figure}[!]
\includegraphics[width=\textwidth]{Plots/f3.png}
\caption{ {{}}
}
\label{fig:scaling-0b}
\end{figure}
\fi
%\begin{figure}[htp]
%\begin{subfigure}{\textwidth}
%\includegraphics[width=\textwidth]{Plots/f3.png}
%\caption{Figure A}
%\end{subfigure}
%\bigskip
%\begin{subfigure}{\textwidth}
%\includegraphics[width=\textwidth]{Plots/f2.png}
%\caption{Figure B}
%\end{subfigure}
%\label{fig:figure3}
%\caption{ }
%\end{figure}

\begin{figure}
\centering
\begin{minipage}{.25\linewidth}
  \includegraphics[width=\linewidth]{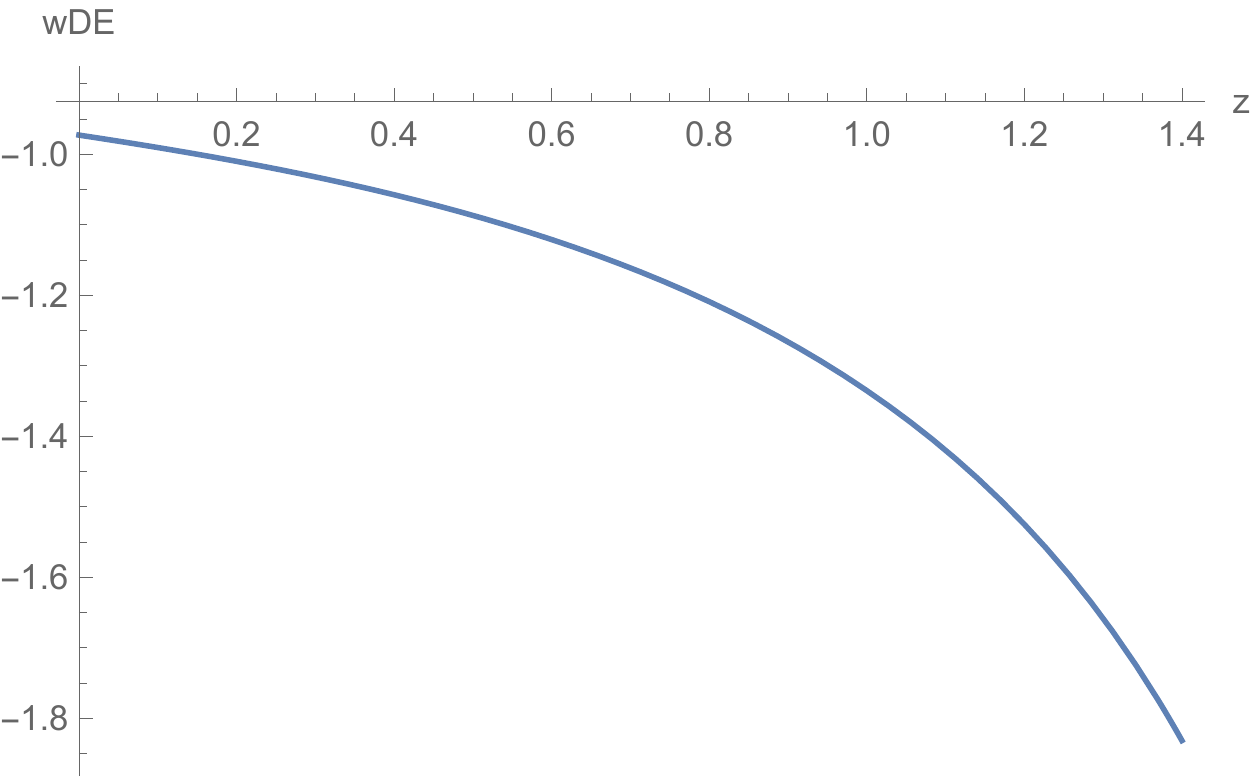}
  %\captionof{figure}{}
  \label{1}
\end{minipage}
\hspace{.05\linewidth}
\begin{minipage}{.25\linewidth}
  \includegraphics[width=\linewidth]{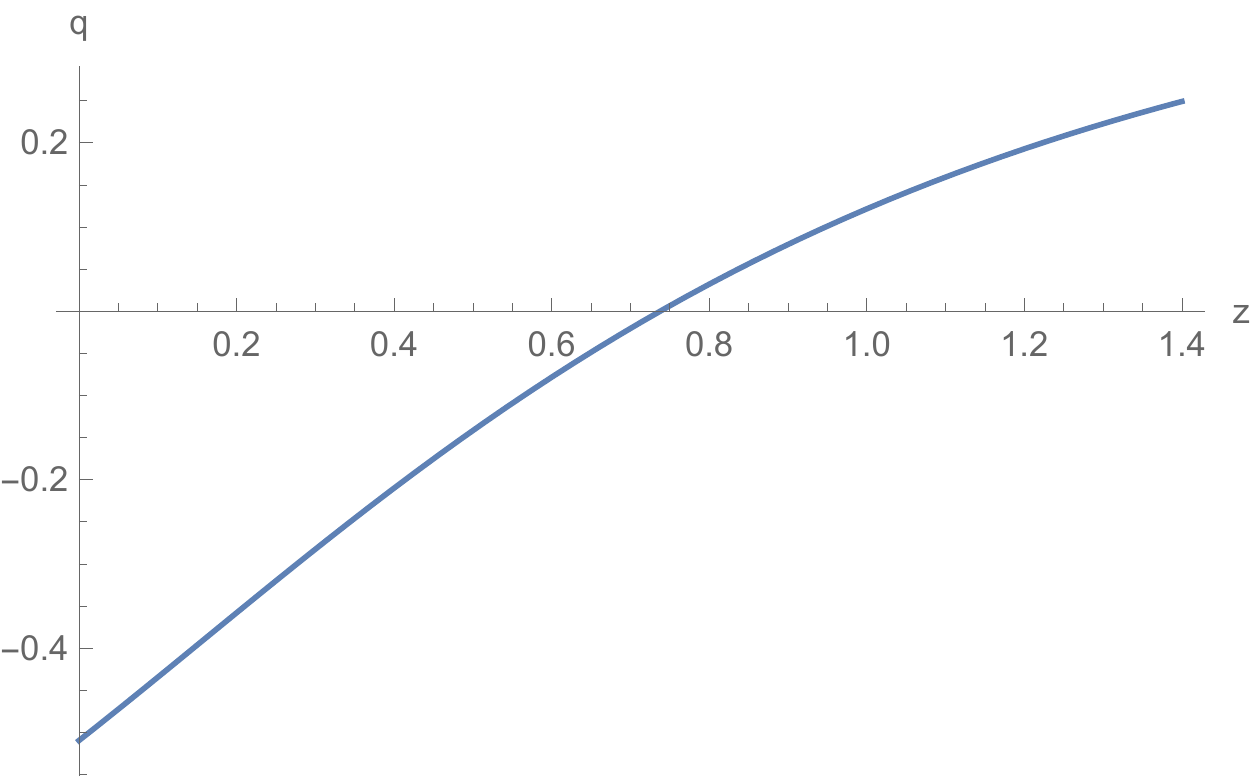}
  %\captionof{figure}{}
  \label{2}
\end{minipage}
\hspace{.05\linewidth}
\begin{minipage}{.25\linewidth}
  \includegraphics[width=\linewidth]{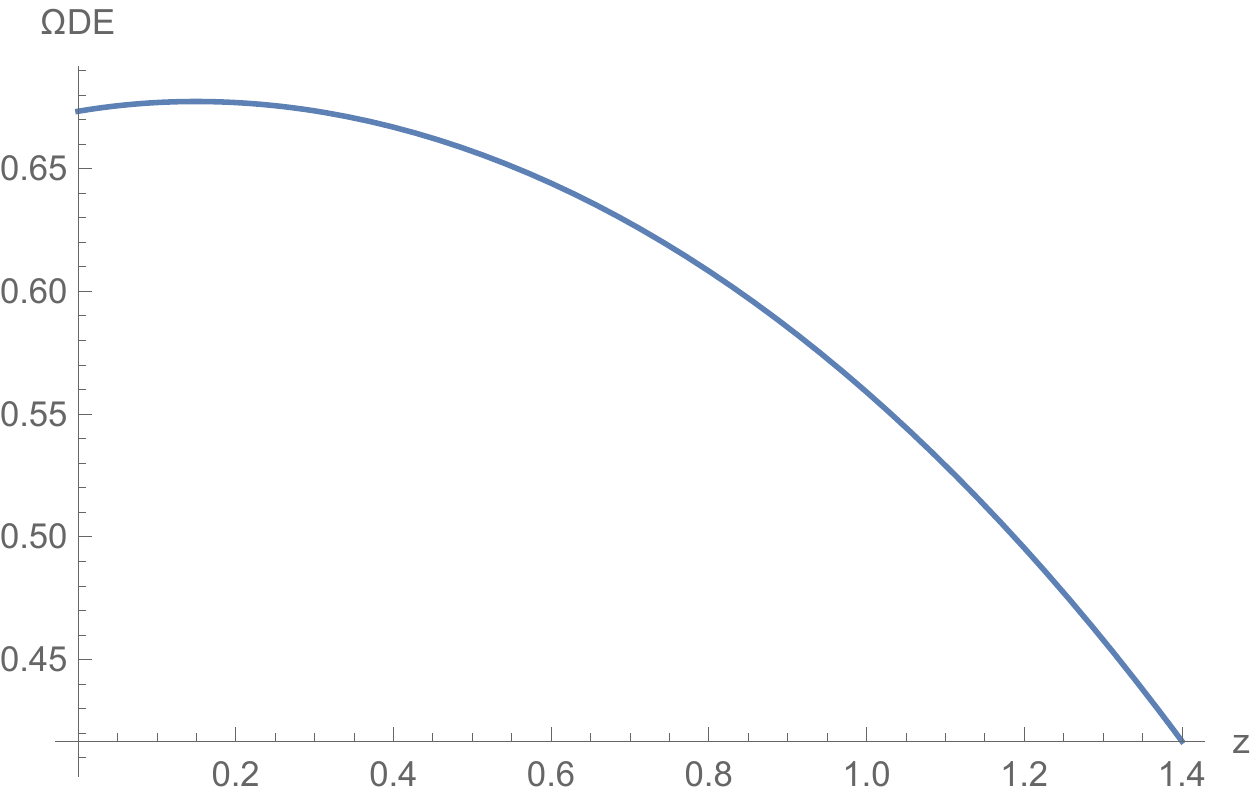}
  %\captionof{figure}{}
  \label{3}
\end{minipage}

\begin{minipage}{.25\linewidth}
  \includegraphics[width=\linewidth]{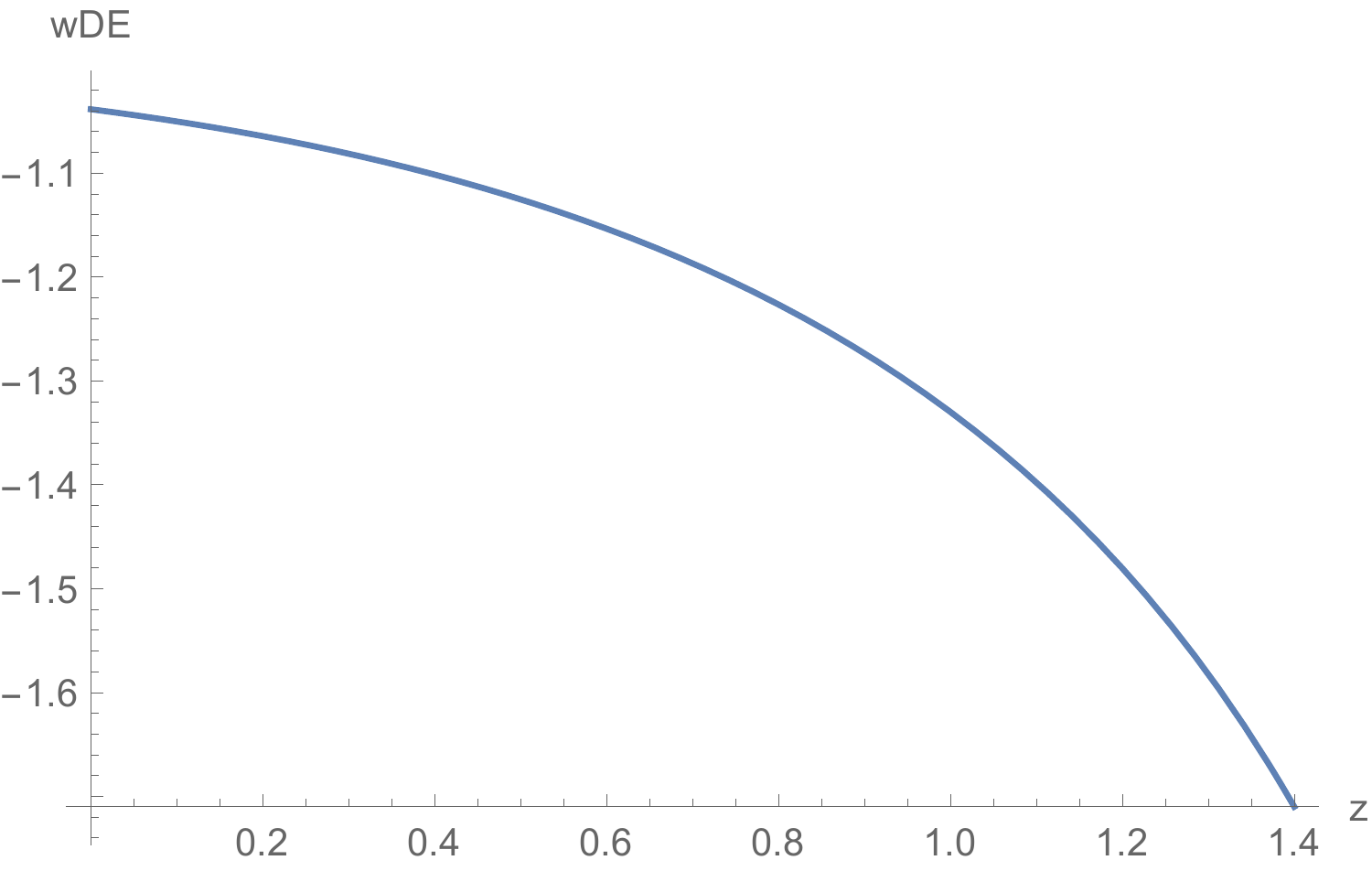}
  %\captionof{figure}{}
  \label{4}
\end{minipage}
\hspace{.05\linewidth}
\begin{minipage}{.25\linewidth}
  \includegraphics[width=\linewidth]{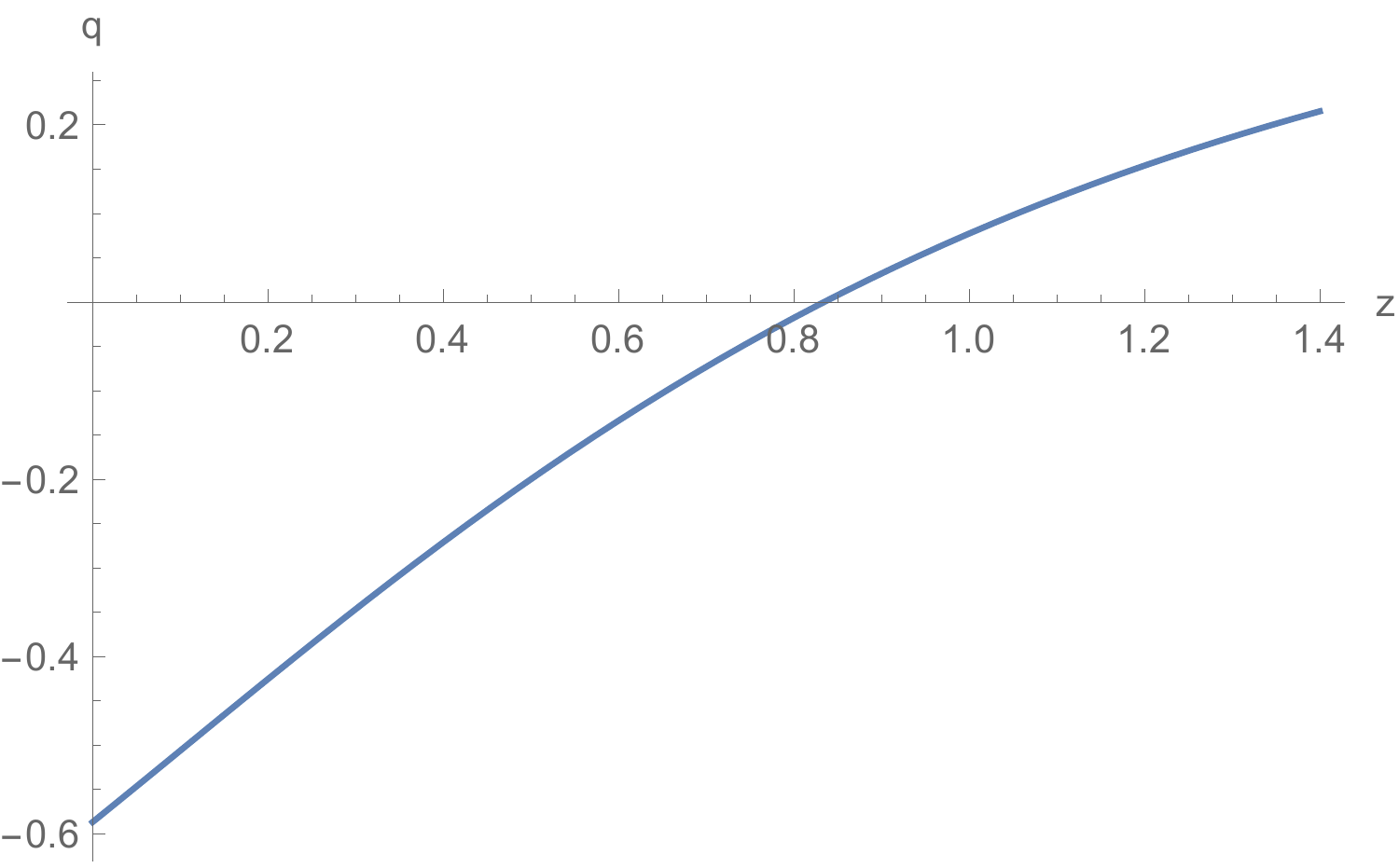}
  %\captionof{figure}{}
  \label{5}
\end{minipage}
\hspace{.05\linewidth}
\begin{minipage}{.25\linewidth}
  \includegraphics[width=\linewidth]{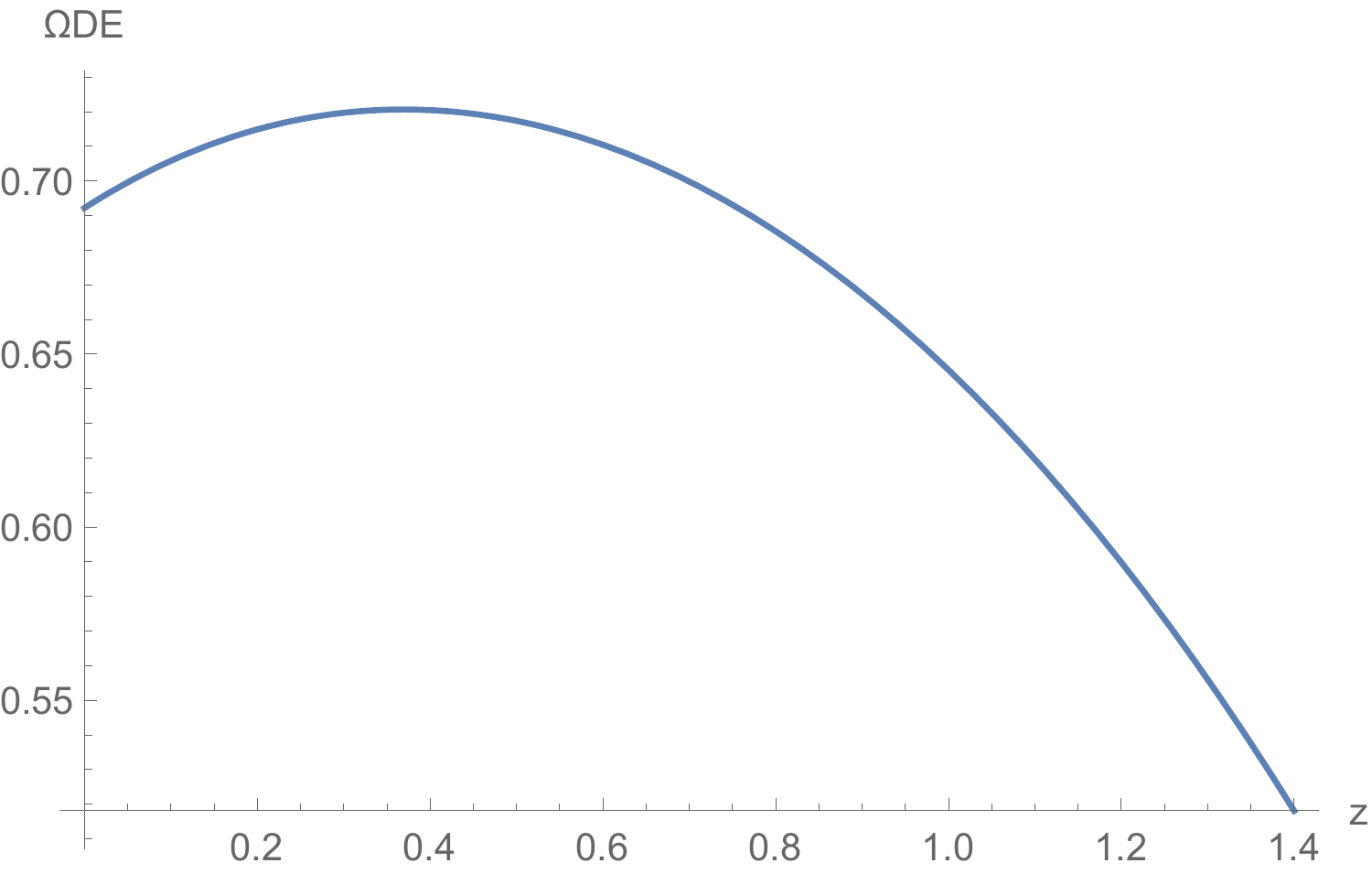}
  %\captionof{figure}{}
  \label{6}
\end{minipage}
\caption{Plots of some representative cosmological parameters. From left to right: equation of state parameter for the effective Dark Energy fluid, $w_{DE}$, deceleration parameter, $q$ and normalized Dark Energy Density, $\Omega_{DE}$ as a function of the redshift $z$, for the two different scaling relations (top and bottom panel), see text. The parameters used for these plots are \textbf{(top panel)} $\Omega_{m0}=0.3, \  \xi = 1, \ \theta_1 = 2.001 , \ \theta_2 = 0.04$, \textbf{(bottom panel)} $\Omega_{m0}=0.3, \  \tilde{\xi} = 10^{-28}, \ \theta_1 = 2.005 , \ \theta_2 = 0.06$.}

\label{fig:figure3}
\end{figure}

\section{Conclusions}
In this work we have extended the  swiss-cheese models presented in \cite{Zarikas:2017gfv,sc3}
to include an explicit running of $G$ according to the IR fixed point hypothesis. 
The presented analysis generalizes the original Einstein-Strauss model by considering an interior vacuole
spacetime described by a RG-improved Schwarzschild spacetime in the presence of a cosmological 
constant and Newton's constant that depend either on the proper distance $D_s$ , or on the
matter density according to a RG trajectory for which both the dimensionless Newton constant
and cosmological constant approach a fixed point at large distances. 

The Darmois matching conditions determine the evolution of the FLRW spacetime outside the
\sch Radius. The RG trajectories emanating from the IR fixed point have been 
parametrized by two critical exponents $\theta_1$ and
$\theta_2$ and by the scaling parameter $\xi$ which relates
the RG cutoff $k$ with a characteristic distance scale. 
It turns out that the present phase of accelerated expansion is the result
of the cumulative antigravity sources contribution within the \sch radius. 
In particular no fine-tuning emerges at late times. Our formalism allows us also to take into account
of the constraints on $\dot{G}/G$. 
Albeit the parameter space is large and a full numerical analysis is beyond the scope of the present work, it was possible to find many 
set parameters which reproduce the current $\Lambda CDM$ standard model. The results are very encouraging, as for both scaling expressions we do obtain reasonable behavior for the cosmological quantities $H(z), q,\,\,w_{DE},$ and $\Omega_{DE}$.

Interesting enough for the interesting cases both $\theta_1$ and $\theta_2$ can be taken as positive and 
this is consistent with the idea that the IR fixed point must be attractive in the
$k\rightarrow 0$ limit.
It would therefore be interesting to further test this  cosmological model against 
the current observational data.

\section{Acknowledgments}
The authors  acknowledge the support of the Faculty Development Competitive Research Grant Program (FDCRGP) Grant No. 110119FD4534. We are grateful to Alessia Platania for important remarks and
comments on the manuscript.
\newpage
\bibliography{ref.bib}

\end{document}